# Essencery – A Tool for Essentializing Software Engineering Practices

Arthur Evensen[1][ORCID to-be-added for final version] Kai-Kristian Kemell[1][Ibid.] Xiaofeng Wang[2][Ibid.] Juhani Risku[1][Ibid.] and Pekka Abrahamsson[1][Ibid.]

[1] University of Jyväskylä, 40014 Jyväskylä, Finland
`arthur.n.evensen@student.jyu.fi`,
`{kai-kristian.o.kemell|juhani.risku|pekka.abrahamsson}@jyu.fi`
[2] Free University of Bozen-Bolzano, 39100 Bozen-Bolzano, Italy
`xiaofeng.wang@unibz.it`

**Abstract.** Software Engineering practitioners work using highly diverse methods and practices, and general theories in software engineering are lacking. One attempt at creating a common ground in the area of software engineering methodologies has been the Essence Theory of Software Engineering, which can be considered a method-agnostic project management tool for software engineering. Essence supports the use of any development practices and provides a framework for building a suitable method for any software engineering context. However, Essence presently suffers from low practitioner adoption that is partially considered to be caused by a lack of proper tooling. In this paper, we present Essencery, a tool for essentializing software engineering methods and practices using the Essence graphical syntax. Essencery aims to facilitate adoption of Essence among potential future users. We present an empirical evaluation of the tool by means of a qualitative, quasi-formal experiment and, based on the experiment, confirm that the tool is easy to use and useful for its intended purpose.

## 1 Introduction

Software Engineering (SE) work out on the field is highly diverse. Practitioners employ a vast variety of methods and practices and typically tailor existing ones to create customized versions to better suit individual organizations [5]. While attempts to introduce universal SE methodologies have been made in the past, the present situation out on the field ultimately underlines the lack of success on this front. In a recent attempt to create a common ground for methods and practices, the SEMAT initiative (semat.org) proposed the *Essence Theory of Software Engineering* (Essence hereafter) [11].

Essence is a modular, method-agnostic framework for SE endeavors that can be tailored to suit any SE context [8]. This modular and extensible nature of Essence, while its primary strength, is also its perhaps greatest weakness. As Essence needs to be tailored for specific contexts to reach its full potential, its adoption is resource-intensive [6]. This resource-intensive adoption process of Essence is, in part, likely to explain its current lack of widespread adoption among practitioners [6, 13].

For development frameworks and methodologies to gain traction among practitioners, tools to support their adoption and use are needed. This is perhaps even more important for Essence given its already resource-intensive adoption. To this end, e.g. Graziotin & Abrahamsson [6] have presented SematAcc to make the Essence kernel easier to utilize. However, further tooling covering other aspects of Essence such as graph-drawing is still needed. The existing Essence Practice Workbench comes with a steep learning curve and is not open source. While pen-and-paper and generic digital drawing tools offer reasonable alternatives, dedicated drawing tool can offer various advantages over general purpose alternatives. E.g. using a dedicated drawing tool, graphs can be far easier to modify and compose, and can offer more for communication purposes.

To further address the present lack of Essence-related tools and to facilitate the adoption of Essence, we present a tool for essentializing software engineering practice, which in practice means drawing graphs using the Essence graphical syntax. In this paper, we develop and evaluate *Essencery – The Essence Practice Editor* through a quasi-formal empirical experiment where IT students (n=16) employ the system to complete a task. Based on the experiment, we evaluate whether Essencery:

1. *Can be used to draw graphs using the Essence graphical syntax, and…*
2. *Is easy to learn and use*

## 2 Theoretical Background

In this section, we discuss the two background theories of this study. The first sub-section discusses Essence in detail while the second sub-section presents the Technology Acceptance Model [3], which is used as the data collection and analysis framework.

### 2.1 The Essence Theory of Software Engineering

Essence is the result of the SEMAT initiative and comprises of a kernel and a language [8]. The kernel is described using a subset of the language, identifying those SE method elements that "are integral to all SE methods". The language contains notational elements that specify elements that are necessarily not part of every SE method, such as specific activities or roles. The kernel contains three views: an alpha view, a competency view and an activity space view. The kernel is extensible, allowing for the creation of customized kernels for e.g. SE endeavors including business analytics.

In practice, SE methods are described using the Essence language, a process that includes relating method-specific elements to the different elements of the Essence kernel. In the simplest case, the kernel itself can serve as a method. Generally, however, the kernel acts as a module in the description of a method and provides the building blocks upon which to begin describing the method.

By being method-agnostic and thus suitable for any SE context, Essence can provide common ground for SE and facilitate creating, adapting and using methods for specific endeavors as opposed to 'one-size-fits-all' "method prisons" [9]. Notably, Essence stands out from all other previously proposed methods and approaches since it has been standardized by the OMG (Object Management Group) [11].

## 2.2 The Technology Acceptance Model

Originally proposed by Davis [3] in 1985, the *Technology Acceptance Model* (TAM) is an influential IS theory intended to explain technology adoption on a general level. The model posits that external variables (e.g. software feature factors) affect the Perceived Usefulness (PU) and Perceived Ease of Use (PEOU) of potential users of a technology. PU and PEoU, then, result in a decision to either accept or reject the technology. In addition, Davis [3] established a link between PEoU and PU: the easier the system is to use, the more useful it often consequently is. Davis [3-4] posits that TAM, seen in Fig 1., is primarily useful for early-stage use testing, specifically in organizational settings.

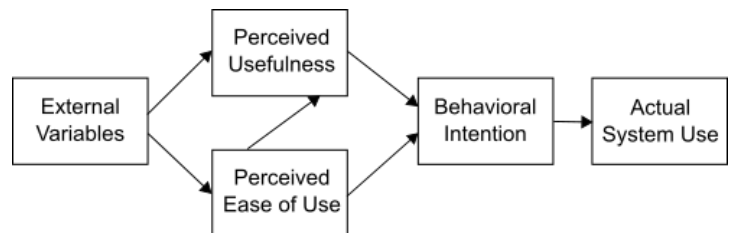

**Fig. 1.** The Revised TAM [4]

The effectiveness of PU and PEOU in the context of technology acceptance has been demonstrated in empirical studies [1]. These two central constructs have remained the same in the sub-sequent versions of TAM (e.g. [4]), and similar constructs have been employed by competing models aiming to predict technology acceptance [e.g. 15]. However, while it has been established that PU and PEOU are effective at predicting technology acceptance, few empirical studies have tried to explain *why* this is the case [1]. In the context of TAM, little effort has gone towards explaining PU and PEOU in terms of system factors [2], although demographic variables have been established to affect perceptions of usefulness and ease of use [15]. In this study, we seek to understand further what system factors affect PU and PEOU in the context of Essencery.

## 3 Essencery – the Essence Practice Editor

*Essencery* is a web application designed primarily PC use, implemented in Ruby on Rails. The tool is intended for drawing graphs using the Essence graphical syntax quickly and easily. The following process for essentializing an SE practice described by Jacobson et al. [7] was used to initially determine the use context of the system:

*1. Identifying the elements – this is primarily identifying a list of elements that make up a practice. The output is essentially a diagram.*
*2. Drafting the relationships between the elements and the outline of each element – At this point, the cards are created.*
*3. Providing further details – Usually, the cards will be supplemented with additional guidelines, hints and tips, examples, and references to other resources, such as articles and books.*

Based on this process description, we decided on the following rough implementation plan: 1) Graph handling support, 2) Card support, 3) Support for adding links and references to outside materials on the cards themselves. At the time of writing of this paper, the first step has been completed. Relations between elements can be visually denoted by no support for practice card creation exists. The tool is thus currently not yet Essence OMG standard [11] compliant and still lacks functionality for kernel binding.

As Essence itself is perceived as resource-intensive to adopt [6], Essencery, seen in Fig. 2 below, was designed to be as simple as possible to not add to this issue. We employed the YAGNI (You Ain't Gonna Need It) principle to implement the essential features while avoiding feature fatigue [14]. Development was carried out inside-out, focusing on application functionality before support functionality such as account systems. In planning the system, we focused on making it as effortless and more powerful than a pen-and-paper session. We implemented a small set of graph handling tools intended to be easy to use while providing all the necessary tools to draw Essence graphs.

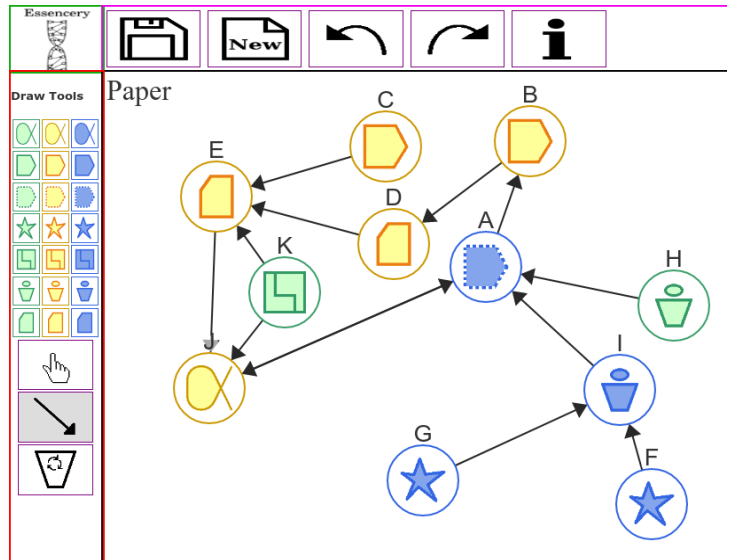

**Fig. 2.** The functional user interface with programmer art

On the left side are the elements of the Essence graphical syntax, with the three colours denoting the three areas of concern of the Essence kernel. Users can point-and-click to place these elements. Placing a node activates the note naming mode. Beneath the buttons for placing elements are four graph drawing tools: one for moving existing nodes (and attached relations) or relations, one for drawing relations, and one for deleting elements, respectively. Not pictured is the text editor tool denoted with a "T..". While the sidebar contains graph handling tools, the top bar contains file handling tools. From the left: save graph, new graph, undo, redo, and return to index of graphs.

Plans for planned graph-drawing features include: re-sizing nodes; improved visuals; selection tools for selecting multiple nodes; kernel binding; an account system; templates; and a sharing system.

## 4 Research Methodology

The empirical data for this study was collected through a quasi-formal experiment. The experiment was designed following the guidelines of Wohlin et al. [16] and Kitchenham [10] for designing qualitative (quasi-)formal experiments in SE. As a framework for collecting and analyzing our data, we employed TAM, and more specifically its two main constructs, Perceived Ease of Use and Perceived Usefulness.

The subjects of the experiment (n=16) were all SE and IS students from the University of Jyväskylä with no prior experience with Essence. The subjects were selected based on their participation in the spring 2018 iteration of a fuzzy front-end course.

In the experiment, the subjects were asked to use the Essence graphical syntax to portray one week of their course project in the fuzzy front-end course by using Essencery to draw a graph of it. The experiment was carried out in sixteen separate sessions, one subject at a time. No introduction to Essencery past its intended use purpose was given. For a brief introduction to Essence, the subjects were provided an A4 sheet describing each graphical symbol in one or two sentences. Data from the experiment were collected using multiple qualitative methods: (1) *the think-aloud protocol*, (2) *participant observation*, and (3) *semi-structured, qualitative interview.*

During the experiment, the participants were instructed to think aloud. The think-aloud protocol employed in the experiment was based on the coaching approach described by Olmsted-Hawala et al. [12]. While active interventions were generally avoided, we instructed the participants to ask for help if needed. If asked for, help was given after a short pause. These pauses were included to ensure that the participants really could not figure out how to achieve what they wanted to do without outside assistance. Occasionally, however, we would briefly intervene to pose questions if a participant was clearly struggling (and not thinking aloud).

Finally, once the subjects announced they had finished drawing their graph, a semi-structured, qualitative interview was conducted with each participant. A set of interview questions, seen in Table 1 below, was used in conjunction with context-specific questions based on e.g. the observation data or their think-aloud statements.

**Table 1.** Semi-structured interview questions

| Question | Aim of the Question |
| --- | --- |
| Do you feel that you managed to draw what you wanted to draw? | To evaluate system PU<br>To determine if task was completed |
| Do you feel that your graph accurately depicts what you did during that week? | To evaluate system PU<br>To determine if task was completed |
| Was the system easy to use? | To evaluate system PEOU |
| Did you have any problems achieving your goal? With the system or otherwise. | To evaluate system PEOU and PU |
| Was there anything you were hoping the software could do but could not do? | To evaluate system PU<br>To map additional user needs |
| In general, how would you like to see the software improved in the future? | To evaluate system PEOU and PU<br>To map additional user needs |

## 5 Data Analysis

The analysis of our data is split into three sub-sections. The first two sub-sections focus on the PEOU and PU of the system, respectively, while in the third sub-section we briefly discuss future adjustments to the system based on the data analysis.

### 5.1 Perceived Ease of Use

The second goal determined for the system was that the system should be "easy to learn and use". In the context of TAM, Davis et al. [4] defines the Perceived Ease of Use as follows: "the degree to which a person believes that using a particular system would be free of effort". The evaluation of the PEOU of Essencery was based primarily on the data gathered through the think-aloud protocol and the interviews.

During the experiment, 15 out of 16 participants experienced no notable difficulties using the system and described the system as being easy to use and largely intuitive in the final interviews. The participants generally encountered only minor issues such as initially attempting to drag-and-drop elements but quickly adjusting to the point-and-click functionality instead. Help requests were rare and were largely related to Essence rather than the system. Occasionally the participants would confirm their own observations through questions or statements (e.g. "I guess the Nordic letters do not work yet").

The one participant that struggled with some of the functionalities of the system expressed no frustration with it and noted that the issues they had would have been quite easily tackled by a brief tutorial to the system, which at the time was not available in the system itself. Many participants also noted that they felt nervous during the experiment and they felt that it made learning to use a new system harder than it perhaps otherwise would have been – even if they felt that it had not been hard as such.

When directly asked about the problems they had faced during the experiment, the participants largely discussed problems with planning out their graphs and Essence rather than problems with Essencery. Some concerns related to Essencery were raised as well, although the participants described them as being minor inconveniences rather than notable problems. For example, many initially assumed that Essencery would be based on drag-and-drop rather than point-and-click functionality. Based on our data, we thus posit that the system is "easy to use and learn".

### 5.2 Perceived Usefulness

The primary goal determined for the system was that it "can be used to draw graphs using the Essence graphical syntax". This goal is associated with the PU of TAM which Davis et al. [4] define to be "the degree to which a person believes that using a particular system would enhance his or her job performance".

When asked whether they had succeeded in drawing what they wanted to draw (i.e. completed their task), all 16 participants agreed. All 16 participants also thought that their graph accurately depicted a week of their course project work. Furthermore, all participants felt that they could have added more details to their graphs if needed. However, they all felt that their graphs were sufficiently detailed to complete the task.

In addition to the self-reported performance of the participants, we utilized their graphs to evaluate task performance in a more objective fashion. Though the participants did not always fully correctly utilize the Essence graphical syntax, which is unsurprising given their lack of history with Essence and the very brief introduction given to it, their graphs did succeed in describing project work in an understandable fashion.

Based on our data, we argue that the system can be considered useful for its intended use purpose of drawing graphs with the Essence graphical syntax. In its current state, most of its usefulness in terms of enhancing job performance stems from the link between the PU and PEOU of the system. I.e. as the PEOU of a system affects its PU [4], due to Essencery being easy to learn and use, it enhances job performance over alternatives such as pen and paper. Additionally, Essencery offers functionalities to reduce the effort needed to modify existing graphs such as the ability to move nodes and their relations at the same time.

### 5.3 Planned Future Improvements to Essencery Based on the Data

As Essencery is still under development, during the interviews we discussed potential improvements with the participants based on their experiences with the system. Excluding features already under development, the following features were included for future implementation based on this data:

- Ability to select multiple elements simultaneously,
- Substituting existing elements with new ones without having to delete them,
- An improved text editor,
- More visual elements such as boxes or different arrow types, and
- Ability to use various standard keyboard shortcuts

## 6  Discussion and Conclusions

In this study, we have developed Essencery, a tool for drawing graphs using the Essence graphical syntax, and conducted early evaluation of it through a quasi-formal experiment using IT students (n=16). Data from the experiment was collected through observation, a think-aloud protocol, and semi-structured interviews. Based on our data, we have demonstrated that it (1) can be used to draw graphs using the Essence graphical syntax, and that (2) it is easy to learn and use.

Essence suffers from a lack of practitioner adoption [13] and extant research [6] has suggested that this is likely to at least partially stem from its difficult and resource-intensive adoption, suggesting tooling as one solution to this issue. We posit that Essencery tackles this problem to an extent. It is useful for practitioners looking to utilize Essence as it provides an easy to learn and use tool for graphs using the graphical syntax of Essence. Though the tool is largely intended to serve as a practical contribution, it can similarly be employed in further studies on Essence (e.g. by having the subjects of a practitioner study employ it to utilize Essence).

Essencery is Open Source and its development continues based on the data gathered for this paper. It is currently usable for graph-drawing, though various future functionalities discussed such as the ability to compose Essence practice cards, as well as Essence OMG standard compliance, are pending implementation. Those potentially interested in the development of Essencery are encouraged to contact the authors.